# VEViD: Vision Enhancement via Virtual diffraction and coherent Detection

Bahram Jalali*, Callen MacPhee
Electrical and Computer Engineering Department, UCLA
*jalali@ucla.edu

**Abstract**
The history of computing started with analog computers consisting of physical devices performing specialized functions such as predicting the trajectory of cannon balls. In modern times, this idea has been extended, for example, to ultrafast nonlinear optics serving as a surrogate analog computer to probe the behavior of complex phenomena such as rogue waves. Here we discuss a new paradigm where physical phenomena coded as an algorithm perform computational imaging tasks. Specifically, diffraction followed by coherent detection, not in its analog realization but when coded as an algorithm, becomes an image enhancement tool. Vision Enhancement via Virtual diffraction and coherent Detection (VEViD) introduced here reimagines a digital image as a spatially varying metaphoric light field and then subjects the field to the physical processes akin to diffraction and coherent detection. The term "Virtual" captures the deviation from the physical world. The light field is pixelated and the propagation imparts a phase with an arbitrary dependence on frequency which can be different from the quadratic behavior of physical diffraction. Temporal frequencies exist in three bands corresponding to the RGB color channels of a digital image. The phase of the output, not the intensity, represents the output image. VEViD is a high-performance low-light-level and color enhancement tool that emerges from this paradigm. The algorithm is interpretable and computationally efficient. We demonstrate image enhancement of 4k video at 200 frames per second and show the utility of this physical algorithm in improving the accuracy of object detection by neural networks without having to retrain model for low-light conditions. The application of VEViD to color enhancement is also demonstrated.

## 1. Introduction

For two thousand years, the Antikythera mechanism lay quietly in the Mediterranean Sea, a timestamp of one of humanity's first known attempts at artificial computing. It is theorized that the machine could calculate the positions of the sun and moon as a function of date and time [1]. Since then several other generations of computing machines were imagined and built, typically with the same continuous state space of this ancient device. Invented in 1206, Castle Clock was a hydro-powered astronomical clock that was the first programmable analog computer [2]. Later, the industrial revolution saw the creation of analog machines that solve differential equations and calculate firing angles of artillery shells [3]. These devices perform a computational task by mapping it into a proxy mechanism that mimics the problem of interest. In this context, optics offers a unique platform for analog computing and realization of physical co-processors for the acceleration of scientific computing [3] such as emulation of Rogue Waves – a stochastically-driven nonlinear phenomenon [4,5]. While analog computers utilize varying degrees of physical abstraction to model the actual system, there remains an underlying continuous space mapping between the states of the machine and the states of the system modeled.

With the advent of much more predictable and governable digital devices, this mapping is violated, resulting in general-purpose computers that are tremendously successful in following any instructions coded in software. Given their theoretical and empirical performance bottleneck manifested in power dissipation and latency, the lure of faster, more efficient analog mappings for niche applications remains. Here we describe such a mapping, namely in the field of low-light image enhancement.

When captured in low-light conditions, digital images often incur undesirable visual qualities such as low-contrast, feature loss, and poor signal to noise ratio. The goal of low-light image enhancement is the abatement of these qualities for two purposes: increased visual quality for human perception and increased accuracy of machine learning. In the former, real-time processing can serve as a boon for convenient viewing, but in the latter, it serves as a requirement for emerging applications such as autonomous vehicles and security. Furthermore, video capture entails a fundamental tradeoff between light sensitivity, which is proportional to exposure time, and frame rate. This obviates the increase in exposure time as a meaningful solution to improving the image quality at low light levels because that would sacrifice the frame rate. In other cases, such as that of live-cell tracking in biology, image enhancement is crucial as low light conditions are necessary to avoid phototoxicity (cell death caused by exposure to light).

Considering the present computational landscape and constraints described above, we introduce a physics-inspired, real-time low-light image enhancement algorithm with a theoretical mapping to the physics occurring in natural systems in analog domain. We show this algorithm has exceptional performance in terms of image quality and computational speed.

## 1.1 Prior Work on Low-light Level Enhancement

There has been a great deal of progress in the task of low-light image enhancement in recent years, primarily due to the adoption of powerful machine learning approaches. We therefore split our brief discussion of prior work on low-light level enhancement into classical algorithms that are deterministic and machine learning approaches which are data-driven.

### Classical Algorithms

The field of low-light image enhancement has a very diverse solution set, with several classical algorithms of varying complexity and performance. While the field still lacks a unifying quantitative theory, Retinex theory has arisen as one of the mainstay concepts in classical approaches. Stemming from concepts in human perception theory concerning decomposition of an image into an illumination and a reflectance constituent, Retinex based approaches account for a large portion of low-light image enhancement techniques [6]. LIME [7] is one such algorithm that utilizes optimized Retinex theory for illumination map generation for high-quality enhancement. Among classical algorithms, it shows very high performance over a large range of lighting conditions [6]. Similarly, histogram equalization [8] methods are a widely used alternative that create an expanded, more uniform histogram for contrast enhancement and increased dynamic range, yet these methods often suffer from color distortion and other artifacts [9]. To diminish these qualities, local and adaptive histogram equalization techniques have been proposed such as CLAHE [9]. Several other classes of traditional algorithms include frequency-based, defogging, and image fusion methods that are used in High Dynamic Range (HDR) techniques.

### Deep Learning Approaches

The proliferation of deep learning algorithms in the last decade has touched many different fields, and image enhancement is no exception. The preponderance of novel algorithms within the field have been data-driven. On the side of supervised learning, one of the first deep learning based approaches, LLNet [10], gave rise to many other autoencoder based designs. Other networks, like MBLLEN [11], EEMEFN [12], TBEFN [13], all make use of similar ground truth datasets for training. In comparison, networks such as Retinex-Net are built upon the theoretical underpinnings of Retinex's human perception theory and therefore are more interpretable.

All these approaches demonstrate high-performance in target lighting conditions, but typically they have difficulty generalizing to greater domains not covered within training data. Other neural network approaches utilize unsupervised learning in the form of generative models such as EnlightenGAN [14]. Lastly, zero-shot techniques that do not require labeled data, such as Zero-DCE [15], have shown good image quality and fast inference speeds. In Zero-DCE, a group of equalizing s-curves are generated at inference time. These curves are learned through a training process that utilizes a set of custom no-reference loss functions that compute several enhancement characteristics such as exposure error and spatial consistency. While this approach needs no ground truth (labeled training data) it still requires training time and diverse image data. Owing to its small network size, however, the network has fast inference time, making it a candidate for real-time image enhancement at certain resolutions.

While these blackbox machine learning models have been revolutionary, they ultimately are restricted by the accuracy of their loss functions in the absence of labeled reference data. As low-light image enhancement still lacks a rigorous quantitative loss function that accurately reflects human perception, these approaches don't perform well when such heuristic metrics fail to correctly define the enhancement as it would be perceived by the user. In other words, the algorithms may produce images that satisfy the minimum loss function requirement but are not perceived as good images by a human viewer.

In this paper we introduce a new low-light level enhancement computer vision algorithms that is derived from the processes of propagation and detection of light. The algorithm emulates the propagation of light through a physical medium with engineered diffractive properties followed by coherent detection. Unlike traditional algorithms that are a sequence of hand-crafted empirical rules or learning based method that are trained and lack interpretability, our physics-inspired approach leverages a law of nature as a blueprint for crafting an algorithm. Such algorithm can, in principle, be implemented in an analog physical device for fast and efficient computation.

## 2. Vision Enhancement via Virtual diffraction and coherent Detection (VEViD)

### 2.1 Physics Framework

Ubiquitous in nature as well as in optical imaging systems, electromagnetic diffraction is a process in which light acquires a frequency-dependent phase upon propagation. This spectral phase is a quadratic function of the spatial frequencies of the light field. While the human eye and common image sensors respond to the power in the light, instruments can work with both the intensity and phase of light, with the latter being measured through coherent detection.

Vision Enhancement via Virtual Diffraction (VEViD) introduced here reimagines a digital image as a spatially varying metaphoric light field. It then subjects the field to the physical processes akin to diffraction and coherent detection. The term "Virtual" captures the deviation from the classical diffraction. The virtual world deviates from the physical world in three aspects. The light field is pixelated, and the propagation imparts a phase with an arbitrary dependence on frequency which can be different from the monotonically-increasing behavior of physical paraxial diffraction. Temporal frequency is restricted to three color bands.

To describe this process, we start with the general solution to the homogeneous electromagnetic wave equation in rectangular coordinate $(x, y, z)$

$$E(x,y,z) = \int_{-\infty}^{+\infty}\int_{-\infty}^{+\infty} \tilde{E}_i(k_x,k_y,0)e^{+jk_z z} e^{j(k_x x + k_y y)} dk_x dk_y \quad 1$$

where $\tilde{E}_i(k_x, k_y, 0)$ is the spatial spectrum of the input field $E_i(x, y, 0)$. Then the Fourier content of the signal after a distance $z$ gains a phase term which can be represented by a spectral phase, $\phi(k_x, k_y)$,

$$\tilde{E}_0(k_x,k_y,z) = \tilde{E}_i(k_x,k_y,0)\, e^{-j\phi(k_x,k_y)} \quad 2$$

and we may rewrite the forward propagated signal subjected to diffractive phase as,

$$E_o(x,y,z) = \text{IFT}\{\tilde{E}_i(k_x,k_y,0)e^{j\phi(k_x,k_y)}\} \quad 3$$

Where IFT refers to the inverse Fourier transform. $E(x, y, z)$ now contains frequency-dependent phase profile that is entirely described by our arbitrary phase propagator $\phi(k_x, k_y)$. The propagation converts a real-valued input $E_i(x, y, 0)$ to a complex function $E_o(x, y, z)$. As described below, we will be interested in the phase of this complex function.

As we are concerned with digital images, we now move from a continuous valued $E(x, y)$ in spatial domain to discrete, meaning pixelated, waveform $E[n, m]$. Similarly in the frequency domain from continuous $(k_x, k_y)$ to discrete momentum $[k_n, k_m]$.

Of primary interest to us is the "lightfield", which we define as the distribution of "field" strength across the two-dimensional landscape of the input signal with the pixel brightness mapped into the metaphoric field strength. The equivalent temporal frequency of the lightfield has three bands corresponding to the three fundamental color channels (RGB). To arrive at the field for color images, we transform our input RGB image into the hsv color space. We will refer to this quantity as $E[n, m; c]$ where $c$ is the index for the color channel To preserve the color integrity, the diffractive transformation operates only on the "v" channel of the image when performing low-light enhancement.

## 2.2 Mathematical Framework

For results that follow, spectral phase filter has a low pass characteristic. A wide range of low pass spectral phase functions can be used. For simplicity a Gaussian function with zero mean and variance $T$ for the frequency dependent phase is considered here,

$$\varphi[k_n, k_m] = \exp\left[-\frac{k_n^2 + k_m^2}{T}\right] \quad 4$$

Resulting in a spectral phase operator,

$$H[k_n, k_m] = S \cdot e^{-i\phi[k_n, k_m]} \quad 5$$

Where S is a model parameter that maps into propagation loss (or gain).

Following the application of the spectral phase and inverse Fourier transform, coherent detection produces the real and imaginary components of the field from which the phase is obtained. The combined processes of diffraction with the low pass spectral phase and coherent detection produces the output of VEViD, $V[n, m]$,

$$V[n, m; c] = angle \langle IFT\left\{e^{-i\phi[k_n, k_m]} \cdot FT\{E[n, m; c]\}\right\}\rangle \quad 6$$

Where FT denotes the Fourier transform, and the angle operator calculates the phase of the complex-valued function of its argument. Previously, other types of spectral phase operations have been exploited in creating edge detection algorithms [16, 17].

## 2.3 Impact of VEViD in Spatial and Frequency Domains

The effect on the spatial domain representation is shown in Figure 1 (top row). The input image is a real valued function. After virtual diffraction, real component is nearly unchanged however the image acquires a significant imaginary component. After phase detection, the image is once again a real valued function but is significantly different from the input.

The effect on the spatial frequency domain is shown in Figure 1 (bottom row). The imaginary portion of the spectrum adopts a central low frequency spike, while the real portion undergoes corresponding attenuation in its low frequency component due to energy conservation.

## 2.4 The VEViD Algorithm

The VEViD algorithm is formally defined in Figure 2. The input image is first converted from RGB to *hsv* color space (not shown). For low light level enhancement the hue channel and saturation channels are not transformed because we seek to retain the original color mapping of the input image. For color enhancement, the *v* and *h* channels are left unchanged and the *s* channel is transformed with VEViD.

A small constant bias term, *b*, is added to the field for the purposes of numerical stabilization and noise reduction. This step is not necessary but it improves the results. The real-valued input image is then transformed into the Fourier domain by FFT and subsequently multiplied elementwise by the complex exponential with an argument which defines the frequency dependent phase. Inverse Fourier transform (IFFT) returns a complex signal in the spatial domain. Mathematically, the inverse tangent operation in phase detection behaves like an activation function. Before computation of phase, the signal is multiplied by a parameter *G* called phase activation gain. The output phase is then normalized to match the image formatting convention [0 to 255]. This output is then injected back into the original image as the new v channel (for low light enhancement) or the s channel (for color enhancement) to obtain the output.

The results shown in Figure 3 demonstrate the quality and generalization of VEViD to several application domains and illumination conditions. We note the ability of VEViD to produced enhanced images with natural colors.

Along with low-light image enhancement, the VEViD transformation is also capable of performing color enhancement for realistic tone matching when applied to the saturation channel of the input image. The process is identical to that of the low-light enhancement procedure described previously with the exception the transform is applied to the *s* channel. The results are shown in Figure 4.

The benefits of having a physical algorithm include low computational burden owing to its simplicity, generalizability to wide range of domains, and the potential for implementation in the analog (physical) domain.

## 2.5 Application to Object Detection via Deep Neural Networks

With the advent of deep learning, computer vision approaches are having spectacular success in applications such as autonomy, manufacturing, and security and defense. On the other hand, these approaches are often unpredictable in complex real-world environments that involve heterogenous data and outliers not represented within the training set. In this section, we demonstrate how pre-processing with VEViD maps image data into a form which improves the accuracy of object detection using off the shelf neural network algorithms without having to retrain them on low light conditions.

The amount of time, memory, and energy required to train a deep neural network, store and recall the millions or billions of model parameters is expensive and is outpacing the growth in semiconductor performance as described by the Moore's Law [18]. We take for example a powerful object detection neural network, YOLO [19], which has tens of million parameters. While the of ability of such networks to learn complex patterns in the data is vast, the efficacy comes down to the vast number of free parameters in the model and the size and richness of the dataset available to fit those parameters. In addition, retraining of a network such as this for domain-specific applications such as low light conditions is burdensome as it requires large new datasets that must be acquired plus the additional computation for training. For an application such as pedestrian detection, it is very important that YOLO generalizes well to low-light conditions, especially in cases such as autonomous driving. Another important application is the security camera. Unless the network is presented with labeled training images captured under low light levels, there is no guarantee that it will function correctly. Preprocessing the image with an algorithm such as VEViD represents a way to easily increase the generalization of these networks to low light level conditions. We see in Figure 5 and Figure 6 the increased performance of the YOLO pretrained model when the image is first processed by VEViD. In each image, VEViD identifies previously obscure features within the mages. In addition, there are several objects missed by the pretrained YOLO network that, after preprocessing with VEViD, are now detected.

Such increased detection accuracy is of vital importance to applications such as autonomous vehicles and security camera systems. In these cases, performance must generalize to night-time environments. In response to this challenge, we have shown that VEViD increases the performance of state-of-the-art neural network inference in such environments. In the next section, we demonstrate VEViD's exceptional computational efficiency – a key attribute for real world applications where low latency is crucial.

## 2.6 Video Rate Object Recognition in Low-Light Conditions

Video applications provide a rich environment for low-light image enhancement due to the tradeoff between image quality and frame rate. Capturing a video at a high frame rate and without blurring requires short integration time (fast shutter). Such low integration time in turn leads to poor image quality at low light levels because fewer photons are collected. Image enhancement can reduce this constraint, with the caveat that in real-time applications, the enhancement procedure itself must be fast enough as to not constrain the frame rate. As we will see below, VEViD performs real-time low-light image enhancement at much higher frame rate than a state-of-the-art neural network technique while producing comparable or better image quality.

Figure 7 shows the runtime vs. the image frame size for VEViD as performed on an NVIDIA GeForce GTX TITAN X Graphic Processing Unit (GPU). Such asynchronous runtimes are measured using specialized timing functions within the PyTorch library (see the Methods section). The VEViD algorithm operates in real-time at a frame rate of 24 FPS past 4K video (8.294,440 Mega pixles). Shown for comparison is the performance of Zero-DCE, a state-of-the-art deep learning algorithm with one of the shortest inference times according to a recent survey [20]. VEViD scales better with frame size than Zero-DCE with the advantage becoming dramatic for 4K frames. This runtime implies that VEViD can be inserted into the camera ISP as a preprocessing step for video applications without sacrificing frame rate. These results show the potential to augment real-time neural network based classification algorithms such as YOLO such that their inference performance generalizes to low illumination conditions with no need for additional training data.

Whereas the full VEViD algorithm enables high-quality enhancement with state-of-the-art computational performance, in the next section we develop a framework for even faster performance through a mathematical approximation. The mathematically accelerated VEViD enables blazing-fast speed with limited penalty in image quality. The resulting equivalent model of VEViD is derived below followed by demonstration of its performance.

## 3. Computational Acceleration

Low latency is a crucial metric for realtime applications including video analytics and broadcast. We are motivated to investigate whether VEViD can be further accelerated through mathematical approximations that reduce the computation time without appreciable sacrifice in image quality. In essence, we are seeking a compact closed-form equivalent model for VEViD. In doing so, we draw inspiration from the field of semiconductor device modeling where complex device physics is approximated as simple, albeit empirical, closed form equations enabling fast simulations of complex circuits consisting of a massive number of those devices [21].

The most time intensive operations in VEViD are the forward and inverse Fourier transforms. We have developed an equivalent formulation that takes place entirely in the spatial domain, this avoids the Fourier transform and significantly improve the runtime of the algorithm.

As shown in Figure 7, this approach, details of which will be released later, provides significant acceleration of the algorithm enabling equivalent of processing of 4K frames at over 200 fps. Furthermore, as shown in Figure 8, the quality of the output compares very favorably with the full numerical version of the algorithm in both the visual quality, and as a preprocessing step for enhancing the accuracy of object detection in low light conditions.

## 4. Conclusion

Physical diffraction and coherent detection can be used as blueprints for the transformation of digital images and videos leading to a new and surprisingly powerful algorithm for low-light and color enhancement. Unlike traditional algorithms that are mostly hand-crafted empirical rules, the VEViD algorithm presented here emulates physical processes and adapts them to the low light level enhancement of digital images. In contrast to deep learning-based approaches, this technique is unique in having its roots in deterministic physics. The algorithms are therefore interpretable and do not require labeled data for training. Although the mapping to physical processes is not precise, in future it may be possible to implement a physical device that executes the algorithm in the analog domain.

We demonstrated low-light enhancement with image quality comparable to the state-of-the-art neural networks but with much lower latency. While the full VEViD algorithm enables high-quality enhancement with high computational speed, we also developed a framework for even faster speed through a mathematical approximation. This enables low-light enhancement on 4k video at 200 frames per second. There are only three model parameters in the full numerical version of the algorithm and only two parameters, $G$ and $b$, in the compact computationally-accelerated version. Although in the present implementation their values are chosen empirically, the same pair of values works over a wide range of application domains and light levels. Future work may consider making these parameters locally adaptive.

Deep neural networks have proven powerful tools for object detection and tracking, and they are the key to autonomous driving and security systems, among others. We showed the utility of VEViD pre-processing to increase the accuracy of object detection by a popular neural network (YOLO). VEViD allows such neural networks that are trained on daylight images to generalize to night-time environments without having to be retrained. The application of VEViD to the color enhancement of digital images is also demonstrated.

## 5. Acknowledgements

This work was partially supported by the Parker Center for Cancer Immunotherapy (PICI), grant no. 20163828, and by the Office of Naval Research (ONR) Multi- disciplinary University Research Initiatives (MURI) program on Optical Computing Award Number N00014-14-1-0505. The authors thanks Yiming Zhou in Jalali Lab for helpful discussions.

## 6. Methods

For all results shown, computations were performed with a NVIDIA GeForce GTX TITAN X Graphic Processing Unit (GPU). The VEViD algorithm was built and run using PyTorch with support for asynchronous computation. Timing metrics were calculated using PyTorch's built-in Event objects. Zero-DCE's runtime results were computed using the code found at https://github.com/Li-Chongyi/Zero-DCE. This code has been made public by the authors of the original paper [15].

Image data form several common low-light image enhancement datasets, as images captured by us were used. Figures 1, 3 and 5 are from [22]. The IR camera security image (Figure 6) is from [23]. Images in Figure 4 were taken with an iPhone 12 Pro Max, except the lighthouse image which is from [24]. The mage in Figure 8 is from a stock photo website [25].

The YOLOv3 object detection algorithm is used for benchmarking AI performance and is built using PyTorch with pretrained weights [26].

## 7. Taxonomy

| Symbol | Definition |
|---|---|
| $c$ | Color Channel |
| $E(x, y; c)$ | "lightfield" in continuous spatial coordinates |
| $\tilde{E}(k_x, k_y; c)$ | "lightfield" in continuous spatial-frequency coordinates |
| $[n, m]$ | Spatial discrete (pixelated) coordinates |
| $[k_n, k_m]$ | Spatial-frequency discrete (pixelated) coordinates |
| $E_i[n, m; c]$ | Input Image |
| $E_o[n, m; c]$ | Output after propagation |
| $V$ | Output Image (VEViD Transform) |
| $\varphi[k_n, k_m]$ | Spectral Phase Function |
| $H[k_n, k_m]$ | Propagation Operator, $\exp(-i * \phi[k_n, k_m])$ |
| $Phase\ Variance\ (T)$ | Variance of the Spectral Phase Function $\varphi[k_n, k_m]$ |
| $Phase\ Scale\ (S)$ | Phase strength: $\max_{n,m} \varphi[k_n, k_m]$ |
| $Phase\ Activation\ Gain\ (G)$ | Phase activation gain, $V = \tan^{-1} G \cdot (\frac{Re\{E_o\}}{Im\{E_o\}})$ |
| $b$ | Regularization constant |
| $N\{\}$ | Normalization function |

# 8. References

[1] https://en.wikipedia.org/wiki/Antikythera_mechanism

[2] Donald Routledge Hill, "Mechanical Engineering in the Medieval Near East", Scientific American, May 1991, pp. 64–69 (cf. Donald Routledge Hill, Mechanical Engineering)

[3] Solli, D. R., & Jalali, B. (2015). Analog optical computing. Nature Photonics, 9(11), 704-706.

[4] Solli, D. R., Ropers, C., Koonath, P., & Jalali, B. (2007). Optical rogue waves. Nature, 450(7172), 1054-1057.

[5] Dudley, J. M., Genty, G., Mussot, A., Chabchoub, A., & Dias, F. (2019). Rogue waves and analogies in optics and oceanography. Nature Reviews Physics, 1(11), 675-689.

[6] E. H. Land and J. J. McCann, ''Lightness and Retinex theory,'' J. Opt. Soc. Amer., vol. 61, no. 1, pp. 1–11, Jan. 1971.

[7] X. Guo, Y. Li, and H. Ling, ''LIME: Low-light image enhancement via illumination map estimation,'' IEEE Trans. Image Process., vol. 26, no. 2, pp. 982–993, Feb. 2017.

[8] L. Li, S. Sun, and C. Xia, ''Survey of histogram equalization technology,'' (in Chinese), Comput. Syst. Appl., vol. 23, no. 3, pp. 1–8, 2014.

[9] G. Yadav, S. Maheshwari and A. Agarwal, "Contrast limited adaptive histogram equalization based enhancement for real time video system," 2014 International Conference on Advances in Computing, Communications and Informatics (ICACCI), 2014, pp. 2392-2397, doi: 10.1109/ICACCI.2014.6968381.

[10] Kin Gwn Lore, Adedotun Akintayo, Soumik Sarkar, LLNet: A deep autoencoder approach to natural low-light image enhancement, Pattern Recognition, Volume 61, 2017, Pages 650-662,

[11] Lv, Feifan & Lu, Feng & Wu, Jianhua & Lim, Chongsoon. (2022). MBLLEN: Low-light Image/Video Enhancement Using CNNs.

[12] Zhu, M., Pan, P., Chen, W., & Yang, Y. (2020). EEMEFN: Low-Light Image Enhancement via Edge-Enhanced Multi-Exposure Fusion Network. Proceedings of the AAAI Conference on Artificial Intelligence, 34(07), 13106-13113. https://doi.org/10.1609/aaai.v34i07.7013

[13] K. Lu and L. Zhang, "TBEFN: A Two-Branch Exposure-Fusion Network for Low-Light Image Enhancement," in IEEE Transactions on Multimedia, vol. 23, pp. 4093-4105, 2021, doi: 10.1109/TMM.2020.3037526.

[14] Jiang, Y., Gong, X., Liu, D., Cheng, Y., Fang, C., Shen, X., Yang, J., Zhou, P., & Wang, Z. (2021). Enlightengan: Deep light enhancement without paired supervision. IEEE Transactions on Image Processing, 30, 2340–2349.

[15] Guo, C., Li, C., Guo, J., Loy, C., Hou, J., Kwong, S., & Cong, R. (2020). Zero-reference deep curve estimation for low-light image enhancement. In Proceedings of the IEEE conference on computer vision and pattern recognition (CVPR) (pp. 1780-1789)

[16] M. Asghari and B. Jalali, "Edge detection in digital images using dispersive phase stretch transform," International journal of biomedical imaging, 2015.

[17] M. Suthar and B. Jalali, "Phase-stretch adaptive gradient-field extractor (PAGE)," in Coding Theory, 2020.

[18] Neil C. Thompson et al. (MIT) 2020 https://arxiv.org/pdf/2007.05558.pdf

[19] Joseph Redmon, & Ali Farhadi (2018). YOLOv3: An Incremental Improvement. ArXiv, abs/1804.02767.

[20] Li, C., Guo, C., Han, L. H., Jiang, J., Cheng, M. M., Gu, J., & Loy, C. C. (2021). Low-light image and video enhancement using deep learning: A survey. IEEE Transactions on Pattern Analysis & Machine Intelligence, (01), 1-1.

[21] Jalali, B., "Device Physics & Modeling", In: InP HBTs: Growth, Processing, and Applications, Jalali, B., Pearton, S.J. (Eds.), (0-89006-724-4). PP. 229 - 263 (December 1994) Artech House

[22] Yang, W., Yuan, Y., Ren, W., Liu, J., Scheirer, W., Wang, Z., Zhang, & et al. (2020). Advancing Image Understanding in Poor Visibility Environments: A Collective Benchmark Study. IEEE Transactions on Image Processing, 29, 5737-5752.

[23] Jia, Xinyu, et al. "LLVIP: A visible-infrared paired dataset for low-light vision." Proceedings of the IEEE/CVF International Conference on Computer Vision, 2021.

[24] Ma, K., et al.: perceptual quality assessment for multi-exposure image
fusion. IEEE Trans. Image Process. 24(11), 3345–3356 (2015).

[25] https://www.pexels.com/video/neo-classical-building-in-city-at-night-9935075/.

[26] https://pjreddie.com/media/files/yolov3.weights.

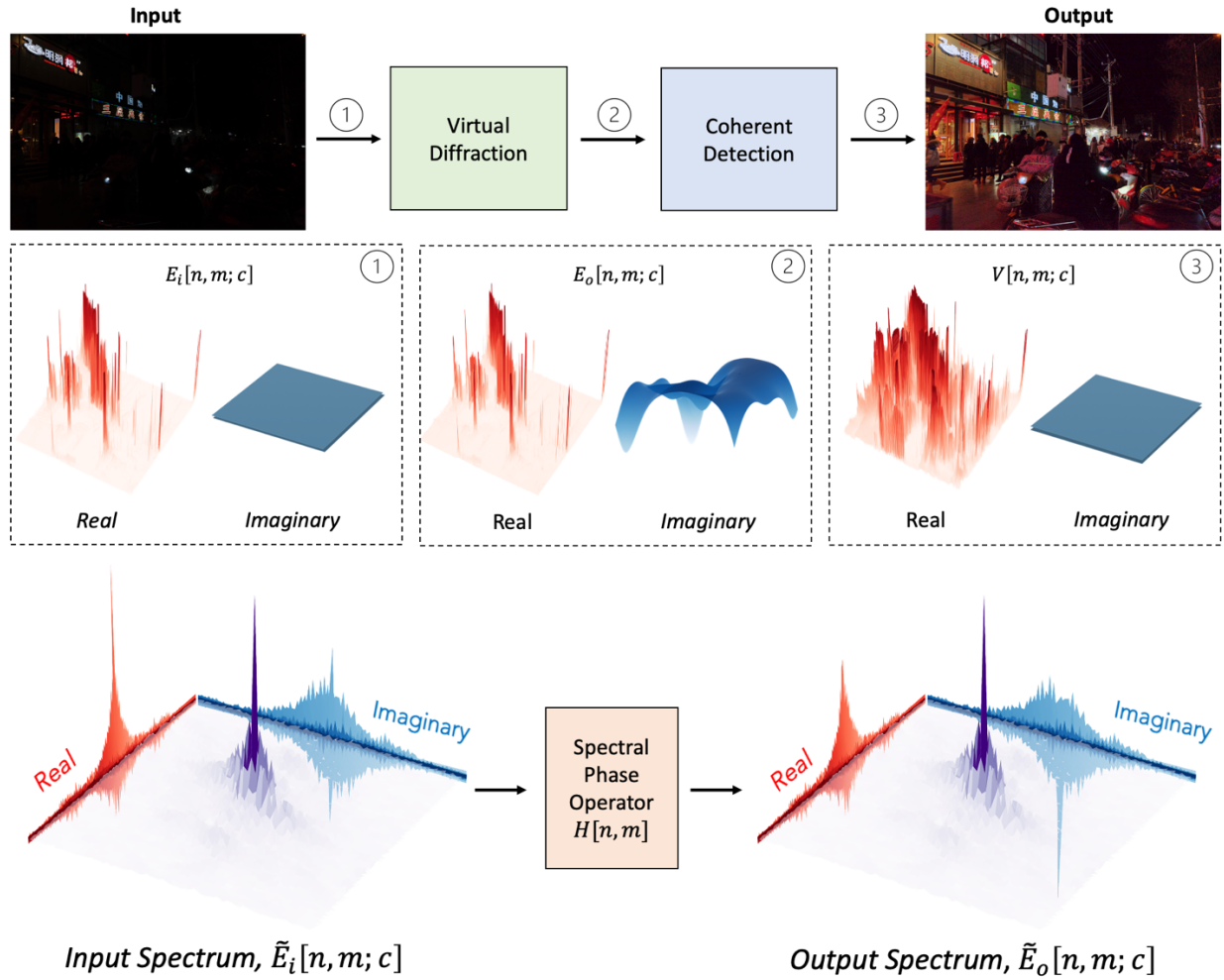

Figure 1: Physical interpretation of the VEViD algorithm showing its impact in spatial domain (top row) and in spectral domain (bottom row). In spatial domain, the real part of the image is nearly unchanged whereas an imaginary part is created after diffraction. This observation supports the mathematical approximation in the latter part of the paper.

**Algorithm 1:** VEViD: Vision Enhancement via Virtual Diffraction

**Data:** $E_i[n, m; c]$, *input image*

**Result:** $V[n, m; c]$, *output image (VEViD transform)*

$\varphi[k_n, k_m] \leftarrow S * \hat{\varphi}[k_n, k_m])$

$H[k_n, k_m] \leftarrow \exp(-i * \varphi[k_n, k_m])$

$E_i[n, m; c] \leftarrow E_i[n, m; c] + b$

$\tilde{E}_i[k_n, k_m; c] \leftarrow FFT\{E_i[n, m; c]\}$

$\tilde{E}_o[k_n, k_m; c] \leftarrow \tilde{E}_i[k_n, k_m; c] * H[k_n, k_m]$

$E_o[n, m; c] \leftarrow IFFT\{\tilde{E}_o[k_n, k_m; c]\}$

$V[n, m; c] \leftarrow \tan^{-1}\left(G * \frac{Im\{E_o[n,m;c]\}}{Re\{E_o[n,m;c]\}}\right)$

$V[n, m; c] \leftarrow N\{V[n, m; c]\}$

*Figure 2: The VEViD algorithm.*

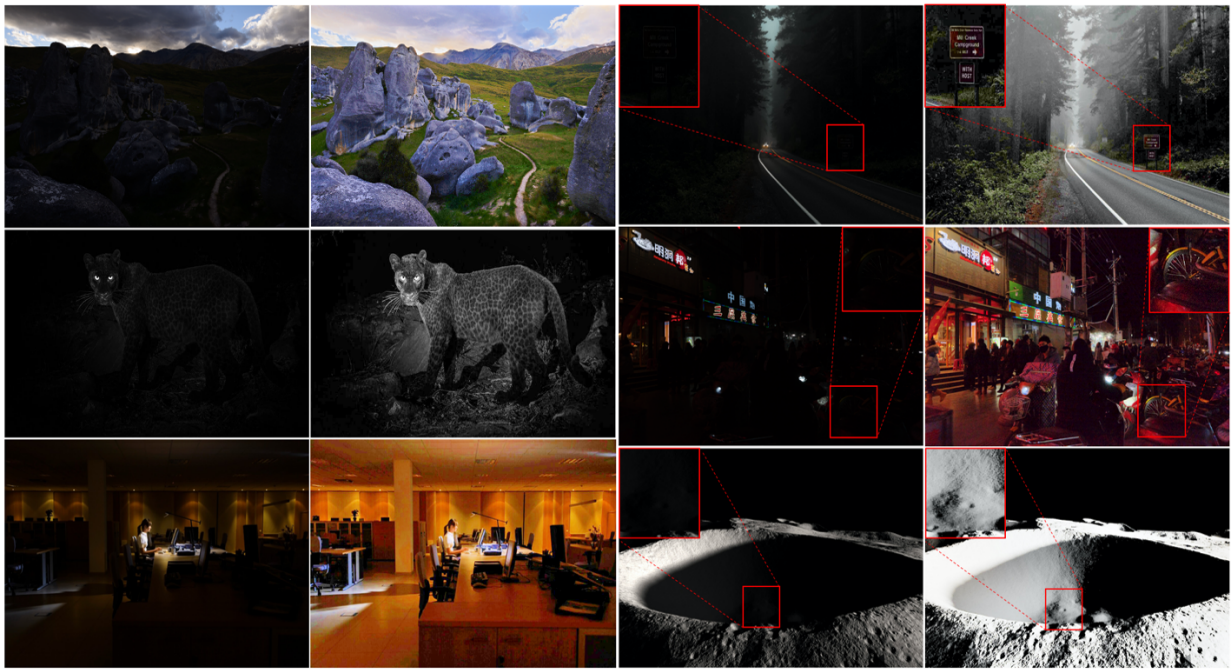

Figure 3: Demonstrating applications of VEViD to various types of images. (Left) Examples of VEViD's low-light enhancement on images captured in low-light conditions. (Right) Examples of VEViD's ability to enhance previously invisible details in low-light images. The same set of model parameter values works over a wide range of images and illuminations.

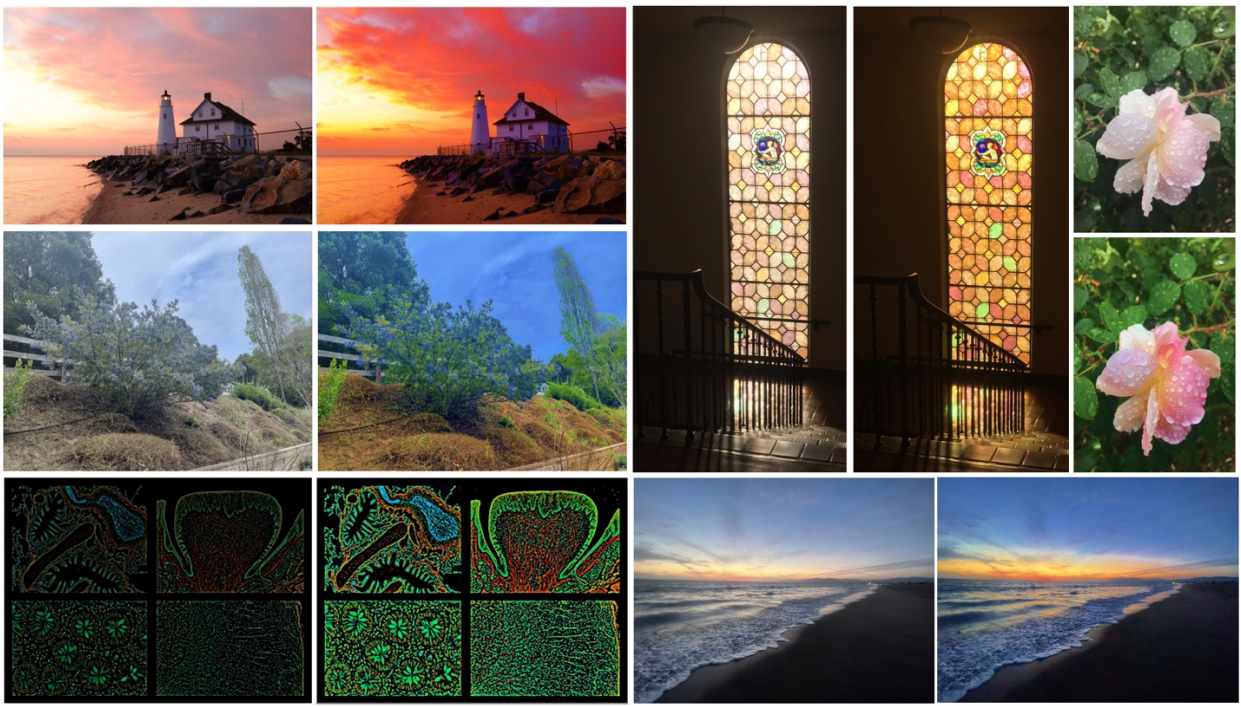

Figure 4: Color enhancement using the VEViD algorithm. Here VEViD operates on the saturation channel of HSV input image. Same set of model parameters are used for all images.

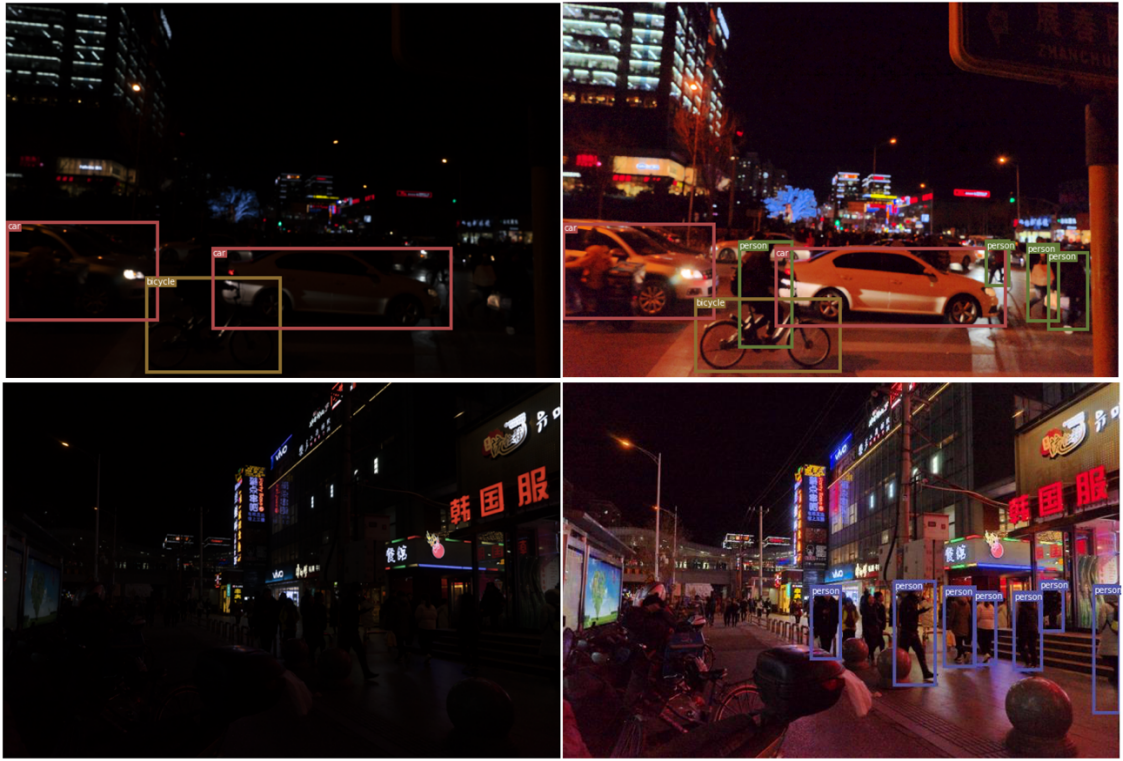

*Figure 5: Impact of VEViD preprocessing on object detection. Examples of object detection using the leading AI neural network (YOLO) on low-light images without (left) and with (right) preprocessing with VEViD. Here, objects recognized by the neural network are shown by bounding boxes. The object detection algorithm is a pretrained YOLO network. Same set of model parameters are used for all images.*

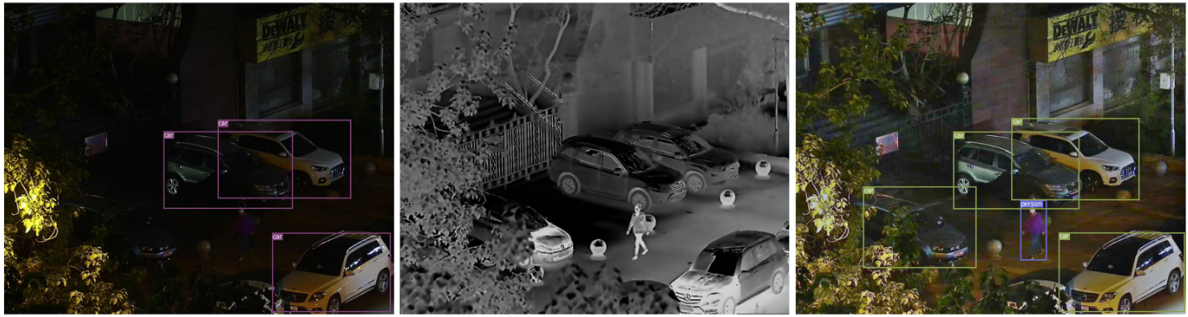

*Figure 6: YOLO performing object tracking on a security camera video on original (left) and VEViD preprocessed images (right). Image from an infrared camera (center) shows that hidden details that are revealed by VEViD match the real scene.*

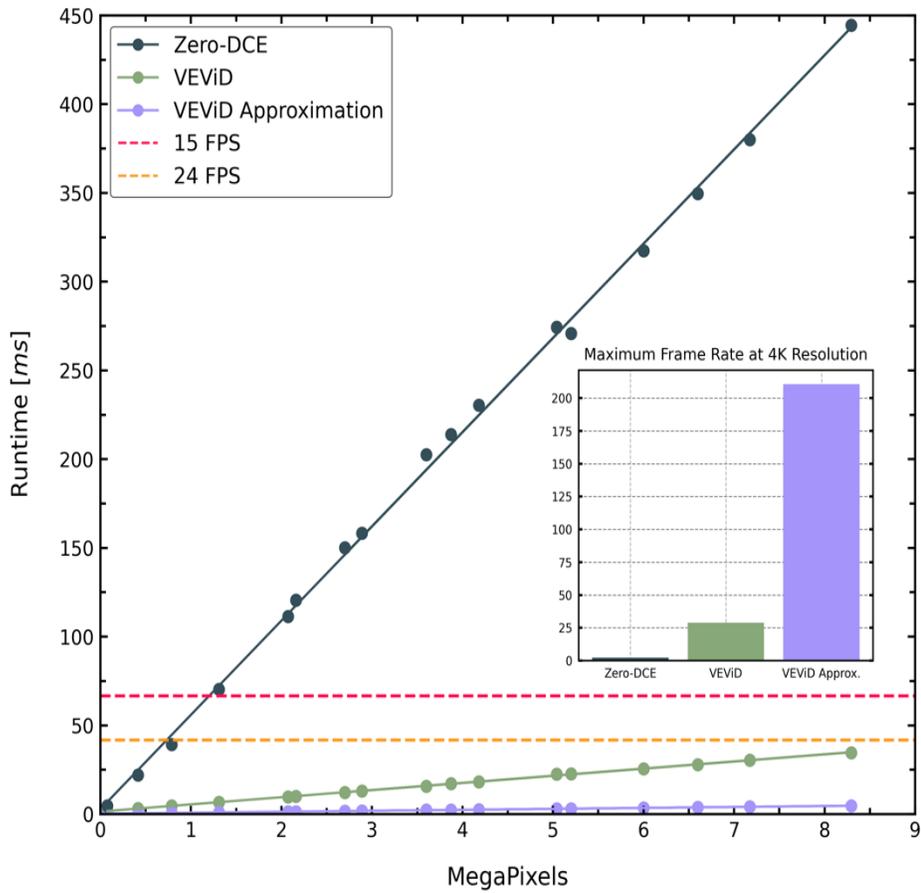

*Figure 7: Comparison of runtime for the VEViD and the Zero-DCE algorithms for a range of image sizes. VEViD is capable of processing at 4K resolutions images (8.294,440 Mega pixels) at 24 FPS and scales much better with frame size. The VEViD approximation is even faster with capability of processing 4K resolution frames at more than 200 FPS. The approximation is a closed-form equivalent-model of the numerical VEViD algorithm that avoids the Fourier transform operations.*

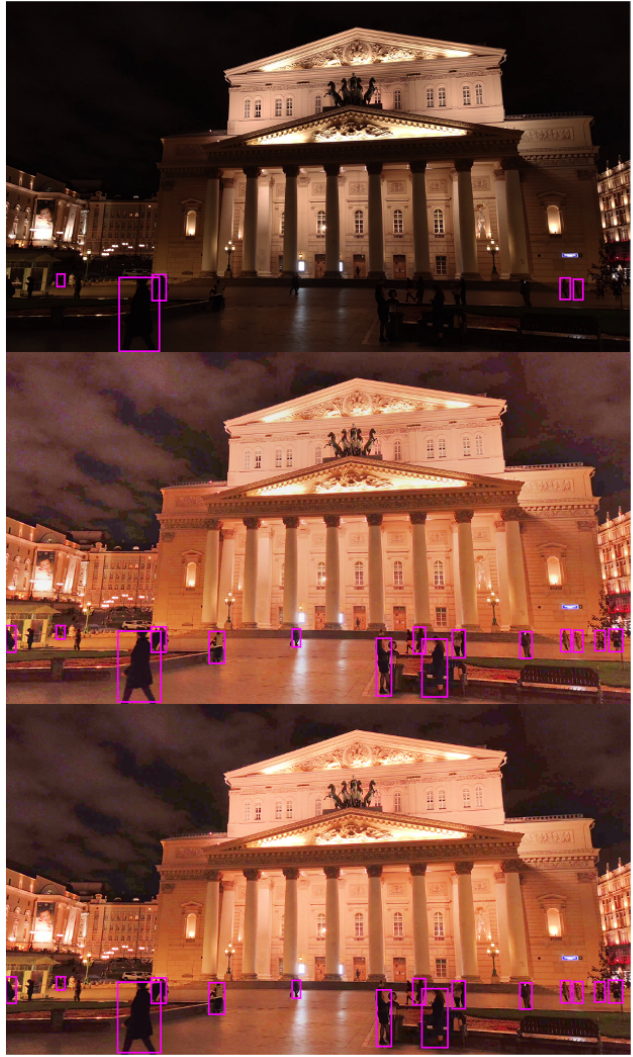

*Figure 8: Impact of preprocessing with VEViD on enhancement of object detection by a neural network (YOLO). When applied to the original image, YOLO identifies 5 objects. After preprocessing by VEViD, the same YOLO algorithm detects 15 objects without having to be retrained on low-light images. Middle image is preprocessed by the full VEViD whereas the bottom image is preprocessed with the approximation. The approximation has very similar visual quality but with much lower latency (Figure 7).*